\documentclass[a4paper,11pt]{article}
\usepackage{pos}

\renewcommand{\d}{\ensuremath{\mathrm d} }
\newcommand{\cD}{\ensuremath{\mathcal D} }
\newcommand{\cO}{\ensuremath{\mathcal O} }
\newcommand{\de}{\ensuremath{\delta} }
\newcommand{\vev}[1]{\ensuremath{\left\langle #1 \right\rangle} }
\newcommand{\eq}[1]{Eq.~\ref{#1}}
\newcommand{\fig}[1]{Fig.~\ref{#1}}
\newcommand{\secref}[1]{Section~\ref{#1}}
\newcommand{\refcite}[1]{Ref.~\cite{#1}}
\title{Advances in using density of states for large-$N$ Yang--Mills}
\ShortTitle{Density of states for large-$N$ Yang--Mills}

\author*{Felix Springer}
\author{David Schaich}
\onbehalf{\\ For the Lattice Strong Dynamics Collaboration}

\affiliation{Department of Mathematical Sciences, University of Liverpool, Liverpool L69 7ZL, United Kingdom}

\emailAdd{felix.springer@liverpool.ac.uk}
\emailAdd{david.schaich@liverpool.ac.uk}

\abstract{We present work in progress using the Logarithmic Linear Relaxation (LLR) density of states algorithm to analyse first-order phase transitions in pure-gauge SU($N$) Yang--Mills theories, focusing on $N = 4$ and 6.  By using the LLR algorithm we aim to avoid super-critical slowing down at such transitions.  Motivation for this study comes from composite dark matter models, which may feature a first-order confinement transition in the early Universe that would produce a background of gravitational waves.  Improving our understanding of these phase transitions will help probe these models using observations from future gravitational-wave observatories.  In addition to the confinement transition, we also analyze bulk phase transitions of the lattice theories, which feature much larger latent heat.}

\FullConference{%
The 39th International Symposium on Lattice Field Theory,\\
8th-13th August, 2022,\\
Rheinische Friedrich-Wilhelms-Universität Bonn, Bonn, Germany
}

\begin{document}
\maketitle

\section{Introduction}
\label{intro}
While standard Markov-chain Monte Carlo importance sampling techniques are an excellent tool for numerous problems in physics, they are fundamentally ill-suited for a range of interesting applications. One of these problems is the low probability of importance sampling algorithms to tunnel between the two co-existing phases at a first-order phase transition, especially on large lattice volumes approaching the thermodynamic limit~\cite{Borsanyi:2022xml, Langfeld:2022uda}. This case of super-critical slowing down can be avoided by using an alternative density of states approach.

There is currently a great deal of interest in first-order phase transitions, due to the possibility that such phenomena in the early Universe could produce an observable background of gravitational waves --- see \refcite{Caprini:2019egz} and references therein.
Our work is motivated by potential gravitational waves from first-order confinement transitions in composite dark matter models, which have also been considered by Refs.~\cite{LatticeStrongDynamics:2020jwi, Huang:2020mso, Kang:2021epo}.
In particular, we are interested in the Stealth Dark Matter model proposed by the Lattice Strong Dynamics Collaboration~\cite{Appelquist:2015yfa, Appelquist:2015zfa, LatticeStrongDynamics:2020jwi}. This is an SU(4) gauge theory coupled to four fermions that transform in the fundamental representation of the gauge group. While these fundamental fermions are electrically charged, they confine to produce a spin-zero, electroweak-singlet `dark baryon'.

Alongside other convenient properties, like a natural way to explain the stability and mass of the dark matter candidate, the symmetries of this model allow it to avoid current direct-detection constraints for heavy dark matter masses $M_{\text{DM}} \gtrsim 1$~TeV~\cite{Appelquist:2015zfa, Appelquist:2015yfa}.
Ongoing research~\cite{LatticeStrongDynamics:2020jwi, Huang:2020mso, Kang:2021epo} investigates if the gravitational waves produced by such composite dark matter models could be detected by future gravitational-wave observatories such as LISA~\cite{Caprini:2019egz}. This would open up a new way to probe these models, which could be especially valuable because collider searches for Stealth Dark Matter pose considerable challenges at heavy masses~\cite{Kribs:2018ilo, Butterworth:2021jto}.

Here we present our progress in employing a particular density of states technique, the Logarithmic Linear Relaxation (LLR) algorithm~\cite{Langfeld:2012ah, Langfeld:2015fua}, to explore the phase transitions of pure-gauge SU($N$) Yang--Mills theories. These theories are interesting for multiple reasons.  First, they are purely bosonic, allowing us to avoid the challenges of applying LLR to systems with dynamical fermions~\cite{Korner:2020vjw}.  In particular, the SU(4) case can be considered the `quenched' limit of the Stealth Dark Matter model, corresponding to infinitely heavy fermions.  For $N \geq 3$ the pure-gauge theories feature first-order confinement transitions of the sort we are interested in exploring, which become significantly stronger as $N$ increases, with latent heat scaling $\propto N^2$ for $N > 3$~\cite{Lucini:2012gg}.

Finally, lattice Yang--Mills theories possess additional bulk (zero-temperature) phase transitions at strong coupling.  For $N \leq 4$ these bulk `transitions' are actually continuous crossovers, becoming weakly first order for SU(5) and strongly first order for $N \geq 6$~\cite{Lucini:2005vg}.  Related to the stronger coupling at which they occur, these $N \geq 6$ bulk transitions are much stronger than the confinement transitions that persist in the physical continuum limit.  That is, they feature much larger latent heat, increasing the advantages of the LLR algorithm compared to importance sampling approaches.

These considerations lead us to focus on the cases $N=4$ and 6 in this proceedings.  This choice provides both the connection to Stealth Dark Matter as well as an opportunity to explore the application of the LLR algorithm to first-order bulk and confinement phase transitions.  The ultimate aims of our work include improving our understanding of the large-$N$ scaling of the latent heat and surface tension, which will assist future studies along the lines of \refcite{Huang:2020mso}.  There is independent work underway studying the weaker SU(3) Yang--Mills confinement transition related to quenched QCD~\cite{Mason:2022trc, Mason:2022aka}.

In the following section we give a brief explanation of the LLR method.
Next in \secref{sec-SU(4)} we present our results from our ongoing LLR analyses of the confinement transition and bulk crossover of SU(4) Yang--Mills, updating Refs.~\cite{Springer:2021liy, Springer:2022qos}.
In \secref{sec-SU(6)} we compare our results for the bulk phase transition of pure-gauge SU(6) Yang--Mills against the SU(4) crossover.
Finally we conclude in \secref{sec-outlook} with a discussion of our current results and a brief outlook on our next steps.

\section{Linear Logarithmic Relaxation algorithm}
\label{sec-LLR}
We begin by considering observables of SU($N$) Yang--Mills theories on the lattice,
\begin{align}
  \label{eq:obs}
  \vev{\cO} & = \frac{1}{Z} \int \cD\phi \, \cO(\phi) \, e^{-S[\phi]} &
  Z & = \int \cD \phi \, e^{-S[\phi]},
\end{align}
where $S[\phi]$ is the lattice action and $\phi$ represents the set of field variables attached to each link in the lattice. Standard Monte Carlo techniques approximate these extremely high-dimensional integrals by analyzing only a modest number of field configurations sampled with probability $\propto e^{-S[\phi]}$.

Alternatively, it is also possible to calculate the density of states
\begin{equation}
  \rho(E) = \int \cD \phi \, \delta(S[\phi] - E)
\end{equation}
and reconstruct the observables of interest as
\begin{align}
  \vev{\cO(\beta)} & = \frac{1}{Z(\beta)} \int \d E \, \cO(E) \, \rho(E) \, e^{\beta E} &
  Z(\beta) & = \int \d E \, \rho(E) \, e^{\beta E}.
\end{align}
Note that this reconstruction is a simple one-dimensional integration.
Since $\rho(E)$ is usually not easily accessible, to compute it we employ the LLR algorithm~\cite{Langfeld:2012ah, Langfeld:2015fua}. In a first step we define the reweighted expectation value
\begin{align}
  \vev{\vev{E - E_i}}_{\de}(a) & = \frac{1}{N}\int \cD \phi \, (E-E_i) \, \theta_{E_i,\de} \, e^{-aS[\phi]} = \frac{1}{N} \int_{E_i-\frac{\de}{2}}^{E_i+\frac{\de}{2}} \d E \, (E-E_i) \, \rho(E) \, e^{-aE}, \label{Heavyside} \\
  N & = \int \cD \phi \, \theta_{E_i,\de} \, e^{-aS[\phi]} = \int_{E_i-\frac{\de}{2}}^{E_i+\frac{\de}{2}} \d E \, \rho(E) \, e^{-a E},
\end{align}
where $E_i$ is a fixed energy value, $\theta_{E_i,\de}$ is the modified Heaviside function ($1$ in the interval $E_i \pm \frac{\delta}{2}$ and $0$ everywhere else), and for now `$a$' is just a free parameter not to be confused with the lattice spacing.

Next we set $\vev{\vev{E - E_i}}_{\de}(a)$ to zero and use the trapezium rule as an approximation:
\begin{align}
  \vev{\vev{E - E_i}}_{\de}(a) & = \frac{1}{N} \int_{E_i-\frac{\de}{2}}^{E_i+\frac{\de}{2}} \d E \, (E-E_i) \, \rho(E) \, e^{-aE} \\
  & =\frac{1}{N} \frac{\de}{2}\left( (\frac{\de}{2})e^{-a(E_i+\frac{\delta}{2})}\rho(E_i+\frac{\de}{2})+(-\frac{\de}{2})e^{-a(E_i-\frac{\de}{2})}\rho(E_i-\frac{\de}{2})\right) + \cO(\de^3) = 0. \nonumber
\end{align}
After performing a Taylor series expansion of $e^{\pm a \frac{\de}{2}}$ and $\rho(E_i\pm \de/2)$ and discarding the $\cO(\de^2)$ terms in the limit $\de \rightarrow 0$, we get
\begin{align}
  0 & = \left(\rho(E_i) + \frac{\de}{2} \frac{\d \rho(E)}{\mathrm{d}E}\Bigr|_{E=E_i}\right)\left(1-a\frac{\de}{2}\right) - \left(\rho(E_i)-\frac{\de}{2}\frac{\d \rho(E)}{\d E}\Bigr|_{E=E_i}\right)\left(1+a\frac{\de}{2}\right) \nonumber \\
  & = \left(-\rho(E_i)a + \frac{\d \rho(E)}{\d E}\Bigr|_{E = E_i} - \rho(E_i)a + \frac{\d \rho(E)}{\d E}\Bigr|_{E=E_i}\right)\frac{\de}{2} \\
  \implies a & = \frac{1}{\rho(E_i)}\frac{\d \rho(E)}{\d E}\Bigr|_{E=E_i} = \frac{\d \ln(\rho(E))}{\d E}\Bigr|_{E=E_i}.
\end{align}
This identifies $a(E_i)$ as a linear approximation of the derivative of the logarithm of the density of states $\rho(E_i)$.
This enables us to calculate the density of states $\rho(E)$, with exponential error suppression~\cite{Langfeld:2012ah, Langfeld:2015fua, Langfeld:2016kty}, by performing a numerical integration of our linear approximation $a(E_i)$ over the intervals $E_i \pm \frac{\de}{2}$, and exponentiating the integral.

The Robbins--Monro algorithm iteratively finds the value of $a$ for a given $E_i$ such that $\vev{\vev{E - E_i}}_{\de}(a) = 0$~\cite{Langfeld:2015fua}:
\begin{equation}
  a^{(n+1)}=a^{(n)}+ \frac{12}{\de^2 (n+1)}\vev{\vev{E - E_i}}_{\de}(a^{(n)}).
  \label{eq:robmon}
\end{equation}
This sequence converges to the correct value of the LLR parameter $a=a^{(n+1)}=a^{(n)}$. The Robbins--Monro algorithm needs the value of the reweighted expectation value $\vev{\vev{E - E_i}}_{\de}(a^{(j)})$ at each iteration in the sequence. This quantity is evaluated using standard importance-sampling Monte Carlo techniques, but with the probability weight $e^{-a^{(j)}S}$ rather than the usual $e^{-S}$.

Due to the modified Heaviside function $\theta_{E_i,\de}$ in \eq{Heavyside}, only configurations with an energy inside of $E_i \pm \frac{\delta}{2}$ are accepted in the Monte Carlo updates, causing lower acceptance rates for smaller energy intervals $\delta$. We can replace this hard energy cut-off with a smooth Gaussian window function~\cite{Langfeld:2016kty, Korner:2020vjw} to alleviate this problem:
\begin{equation}
  \label{eq:gauss}
  \vev{\vev{E - E_i}}_{\de}(a) = \frac{1}{N} \int \d E \, (E-E_i) \, \exp\left[-\frac{(E - E_i)^2}{2\de^2}\right] \, \rho(E) \, e^{-aE}.
\end{equation}
The probability weight in the Monte Carlo simulation is now effectively $\exp\left[-\frac{(E - E_i)^2}{2\de^2}\right] e^{-aE}$. Unlike the modified Heaviside function, the Gaussian window function is differentiable, allowing us to use the hybrid Monte Carlo (HMC) algorithm to evaluate the reweighted expectation value.
In our work we test and compare both ways of constraining the energy interval.
% ------------------------------------------------------------------

% ------------------------------------------------------------------
\section{SU(4)}
\label{sec-SU(4)}
Building on our earlier work presented in Refs.~\cite{Springer:2021liy, Springer:2022qos}, we are using the LLR algorithm to analyze lattice SU($N$) Yang--Mills theories defined by the action
\begin{equation}
  \label{eq:wilson}
  S = -\beta \sum_{x,\mu<\nu} \mathrm{Re}\mathrm{Tr}\left(U_{\mu\nu}(x)\right),
\end{equation}
with the plaquette $U_{\mu\nu}(x) = U_{\mu}(x) U_{\nu}(x+\hat{\mu}) U_{\mu}^{\dagger}(x+\hat{\nu}) U_{\nu}^{\dagger}(x)$.
Here $\beta = \frac{2N}{g_0^2}$ with $g_0^2$ the bare gauge coupling, the sum runs over all lattice sites and $U_{\mu}(x)$ is the SU($N$)-valued link variable attached to lattice site $x$ in direction $\hat{\mu}$.
As a starting point for the implementation of the LLR algorithm, we made use of Stefano Piemonte's \texttt{LeonardYM} software.\footnote{\texttt{\href{https://github.com/FelixSpr/LeonardYM}{github.com/FelixSpr/LeonardYM}}}
We are currently developing our own large-$N$ Yang--Mills code based on the MILC software.\footnote{\texttt{\href{https://github.com/daschaich/LargeN-YM/tree/dev}{github.com/daschaich/LargeN-YM}}}

For $N = 4$, we have tested several updating schemes to compute the reweighted expectation value $\vev{\vev{E - E_i}}_{\de}(a^{(n)})$, including overrelaxation updates in the full SU($N$) group~\cite{Creutz:1987xi}, the Metropolis--Rosenbluth--Teller (MRT) algorithm with SU($N$) updates generalized from \refcite{Katznelson:1984kw}, and HMC updates. % Also did over-relaxed quasi-heatbath...
For the overrelaxation and MRT updates we further compare the hard energy cut-off method \eq{Heavyside} with the Gaussian window approach of \eq{eq:gauss}. As mentioned in \secref{sec-LLR}, in the HMC case only the Gaussian window approach is possible.
The results we obtain are in agreement for all five different updating schemes.

\begin{figure}[btp]
  \centering
  \includegraphics[width=6.4cm]{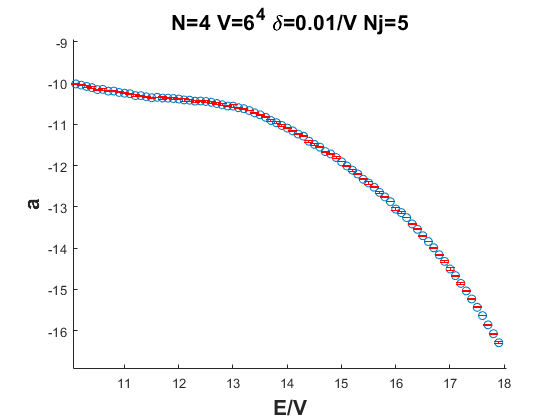}\hfill
  \includegraphics[width=6.4cm]{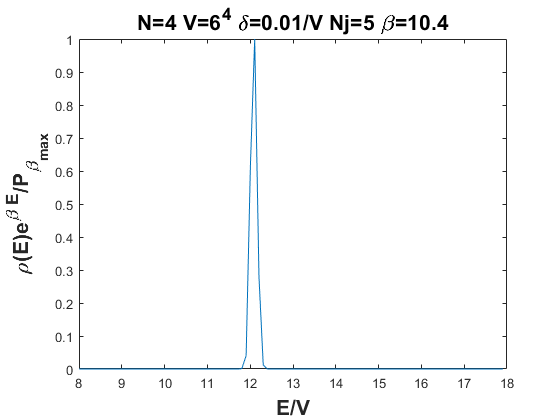}
  \caption{SU(4) results from $6^4$ lattices with an energy interval size of $\de = 0.01/V$.  \textbf{Left:} The LLR parameter $a$, with statistical uncertainties obtained by performing $N_{\text{j}} = 5$ independent runs per interval. \textbf{Right:} The resulting probability density $P_{\beta}(E) = \rho(E) e^{\beta E}$ (omitting uncertainties) at $\beta=10.4$. The single peak structure persists for all $\beta$, signaling that there is no first-order phase transition.}
  \label{fig:avse}
\end{figure}

On an $N_s^3 \times N_t$ lattice, the critical temperature of the first-order SU(4) confinement transition is $T_c = 1 / (a_c N_t)$, corresponding to a critical lattice spacing $a_c$ set by the coupling $\beta_c$.
The continuum limit involves $a_c \to 0$ with $N_t \to \infty$ to keep $T_c$ fixed, implying $\beta_c \to \infty$.
The strong-coupling bulk transition behaves differently, appearing at a smaller $N_t$-independent $\beta_{\text{bulk}}$.
For small $N_t$, $\beta_c$ can approach $\beta_{\text{bulk}}$, causing the confinement transition to be distorted by the nearby bulk transition --- even for $N \leq 4$ where the latter is a continuous crossover.
Based on prior work including Refs.~\cite{Wingate:2000bb, LatticeStrongDynamics:2020jwi}, we consider $N_t \geq 6$ in order to avoid this problem.

We can investigate the bulk transition even with a small, symmetric $6^4$ lattice volume.
Figure~\ref{fig:avse} shows our results for this case over a wide range of energies, using energy interval size $\de = 0.01/V$ and $N_{\text{j}}=5$ independent runs of the Robbins--Monro algorithm for each energy interval.
In order for the probability density
\begin{equation}
  P_{\beta}(E)=\rho(E)\exp(\beta E)=\exp(\int_{-\infty}^{E} \d E^{\prime} a)\exp(\beta E)
\end{equation}
in the right panel of the figure to have the two-peak structure of a first-order transition with co-existence of phases, the LLR parameter $a$ in the left panel must be non-monotonic vs.\ $E / V$.
(See \refcite{Mason:2022aka} for a nice illustration of this.)
Although $a$ decreases less rapidly around the bulk crossover, $E / V \approx 12$, it remains monotonic across all energy intervals.
From this it follows that the probability density only ever features a single peak, which merely broadens around $\beta_{\text{bulk}} \approx 10.4$, consistent with the expected bulk crossover.
Here and throughout this proceedings, we reconstruct the probability density $P_{\beta}(E)$ using both a naive trapezium-rule integration and a more robust polynomial fit technique~\cite{Francesconi:2019nph,Francesconi:2019aet}. In all cases the results from the two techniques are in agreement.

\begin{figure}[btp]
  \centering
  \includegraphics[width=6.5cm]{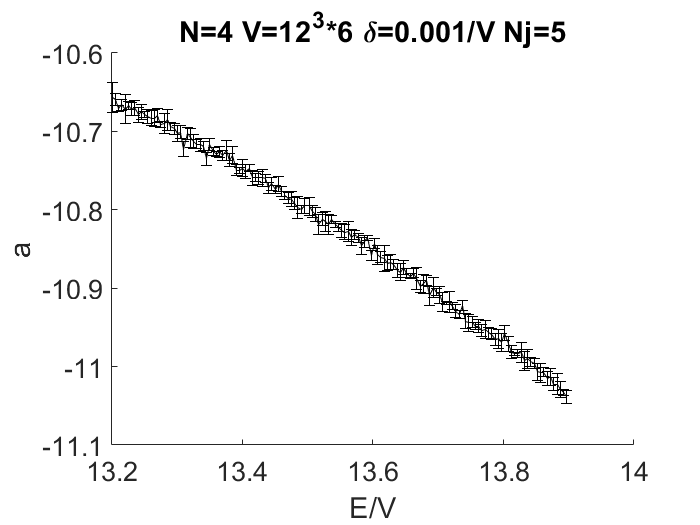}\hfill
  \caption{SU(4) results for $a$ from $V=12^3 \times 6$ lattices with an energy interval size of $\delta = 0.001/V$.  The statistical uncertainties are obtained by performing $N_{\text{j}} = 5$ independent runs per interval. There is no sign of a first-order phase transition, which would correspond to a non-monotonic $a(E)$.}
  \label{fig:avse_conf}
\end{figure}

Now that we have confirmed the crossover nature of the bulk transition, we turn to the physically interesting first-order confinement transition of pure-gauge SU(4). To examine this transition, we have to use a lattice with an aspect ratio $r \equiv N_s / N_t > 1$.
From the previous study \refcite{Wingate:2000bb}, carried out on $N_t = 6$ lattices using importance sampling, we can narrow down the energy range we need to scan to $13.2 < E/V < 13.9$.
Our results for lattice volume $V=12^3 \times 6$, with energy interval size $\de = 0.001/V$ and $N_{\text{j}}=5$ independent runs of the Robbins--Monro algorithm for each energy interval are displayed in \fig{fig:avse_conf}.
We find monotonically decreasing $a(E)$ across all energies, meaning we are so far unable to resolve the expected first-order confinement transition.

\begin{figure}[btp]
  \includegraphics[width=5.8cm]{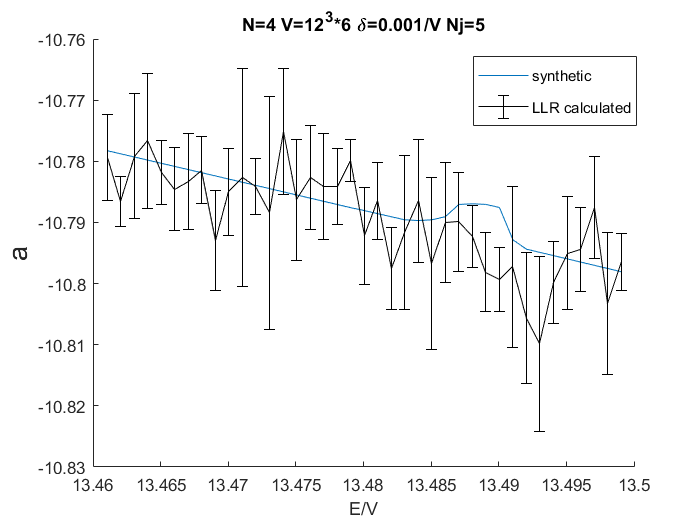}\hfill
  \includegraphics[width=6.0cm]{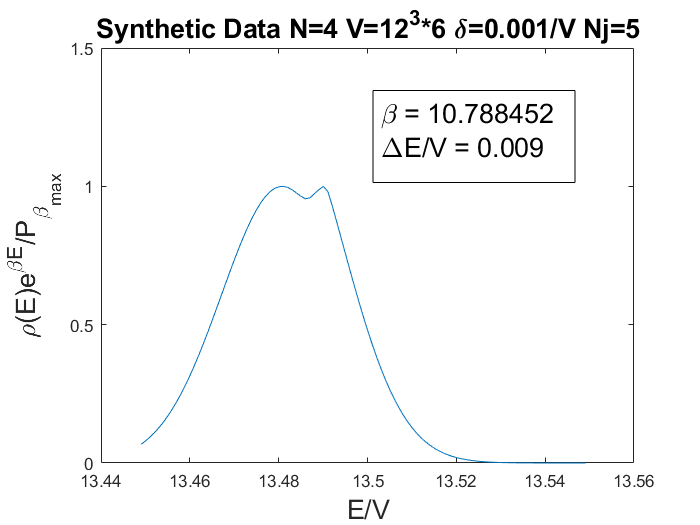}
  \caption{The left plot compares numerical SU(4) data for $a$ (black) with the synthetically generated values (blue line) that would be needed to reconstruct a first-order confinement transition similar to the one in \refcite{Wingate:2000bb}.  Both the synthetic and the measured $a$ use lattice volume $V=12^3 \times 6$.  The right plot shows the probability density $P_{\beta}(E) = \rho(E) e^{\beta E}$ reconstructed from the synthetic $a(E)$ with $\beta \approx 10.8$.  The small double-peak structure illustrates the weakness of this first-order transition.}
  \label{fig:comparison_synth}
\end{figure}

We suspect the reason why we have not yet resolved this phase transition is that it is simply too weak given the small $12^3 \times 6$ lattice volume and our current statistics.
The large-$N$ relation (Eq.~157) in \refcite{Lucini:2012gg} predicts an $N_t = 6$ SU(4) latent heat of only $\Delta E / V \approx 0.004$, which is consistent with \refcite{Wingate:2000bb}.
To explore this, we generated synthetic values of $a$ that would correspond to a first-order transition with a similar $\beta_c \approx 10.8$ and latent heat $\Delta E / V \approx 0.009$. From the comparison of the synthetic data and our actual numerical data in \fig{fig:comparison_synth}, we can see that our statistical uncertainties are too big to detect such a weak first-order phase transition. These considerations give us an estimate for the required improvements in statistics, which may benefit from larger lattice volumes~\cite{Langfeld:2015fua}.
They also motivate analyzing first-order transitions with larger latent heat, in particular strong-coupling bulk transitions for larger $N > 4$.
% ------------------------------------------------------------------

% ------------------------------------------------------------------
\section{SU(6)}
\label{sec-SU(6)}
To investigate the performance of the LLR algorithm for a stronger first-order phase transition, we analyze the bulk transition of SU(6) Yang--Mills theory.
The action (\eq{eq:wilson}), updating schemes, cut-off methods (Heaviside vs.\ Gaussian) and software packages are the same as in the SU(4) case discussed above.

\begin{figure}[btp]
  \centering
  \includegraphics[width=6.5cm]{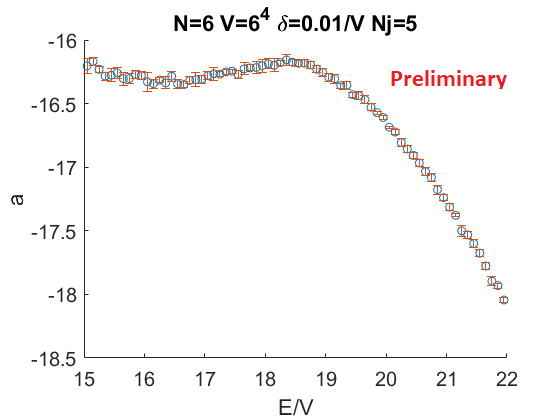}\hfill
  \caption{SU(6) results for $a$ from $V=6^4$ lattices with an energy interval size of $\de = 0.01/V$.  The statistical uncertainties are obtained by performing $N_{\text{j}} = 5$ independent runs per interval.}
  \label{fig:avsenc6}
\end{figure}

\begin{figure}[btp]
  \includegraphics[width=6.4cm]{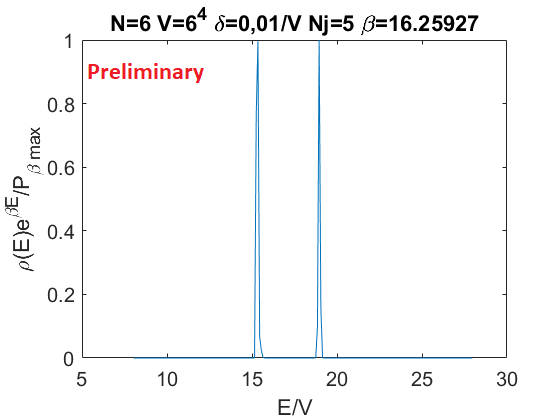}\hfill
  \includegraphics[width=6.4cm]{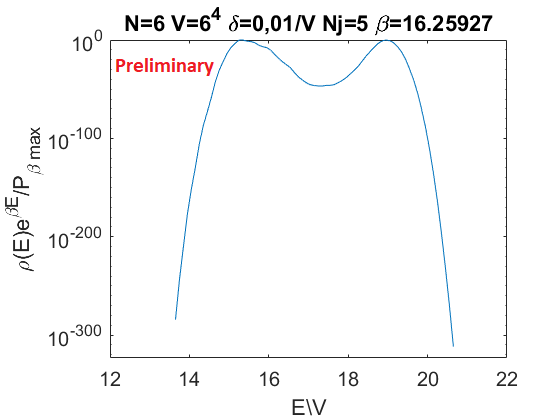}
  \caption{Plot of the probability density $P_{\beta}(E) = \rho(E) e^{\beta E}$ (omitting uncertainties) for pure-gauge SU(6) Yang--Mills at the bulk phase transition ($\beta_{\text{bulk}}=16.25927$), using a lattice of volume $V=6^4$ and an energy interval size $\delta = 0.01/V$. Identical results are plotted with linear (left) and logarithmic (right) y-axes.  The latent heat $\Delta E / V \approx 3.6$ can be directly read off as the distance between the two peaks.}
  \label{fig:nc6probdens}
\end{figure}

In \fig{fig:avsenc6} we show results for the LLR parameter $a$ across a wide range of energies for $6^4$ lattices, an energy interval size $\de = 0.01/V$ and $N_{\text{j}}=5$ independent runs of the Robbins--Monro algorithm for each energy interval. We can clearly see that $a(E)$ is non-monotonic, and increases for $16.5 \lesssim E/V \lesssim 18.5$. This corresponds to the clear double-peak structure plotted in \fig{fig:nc6probdens} (with both linear and logarithmic y-axes) for $\beta_{\text{bulk}} \approx 16.3$.
From \fig{fig:nc6probdens} we can read off a latent heat $\Delta E / V \approx 3.6$, roughly three orders of magnitude larger than the value estimated for the SU(4) confinement transition in \secref{sec-SU(4)}, illustrating the strength of this first-order SU(6) bulk transition.

While such a strong transition would be expected to cause difficulties for traditional importance-sampling studies, it actually simplifies the LLR analyses.
Far less statistical precision is needed to confirm the presence of a first-order transition, precisely determine its critical $\beta_c$, and extract its latent heat and other observables.
A clear next step is to pursue the confinement transition for SU(6) Yang--Mills, and potentially for larger $N$.
Although the latent heat increases $\propto N^2$~\cite{Lucini:2012gg}, for SU(6) this would still lead us to expect a $\Delta E / V \approx 0.004 (6/4)^2 \approx 0.009$ far smaller than that of the bulk transition.
Since computational costs increase $\propto N^3$, it is not a priori clear how practical it will be to resolve large-$N$ transitions using available computing resources. % Not including additional cost increases from need for smaller HMC step sizes

\section{Conclusion and outlook}
\label{sec-outlook}
In this proceedings we presented our progress applying the LLR density of states algorithm to investigate first-order transitions in SU(4) and SU(6) lattice Yang--Mills theories, building on our earlier work in Refs.~\cite{Springer:2021liy, Springer:2022qos}.
Motivation for the pure-gauge theories under consideration comes from our interest a potential first-order dark-sector confinement transition in the early Universe and the stochastic gravitational-wave background it would produce.
Motivation for using the LLR algorithm comes from a desire to avoid the super-critical slowing down occurring for standard Markov-chain importance-sampling techniques at first-order transitions.
Once such first-order transitions are resolved, this density of states approach makes it straightforward to calculate quantities such as the latent heat and surface tension, both of which are needed to predict the spectrum of gravitational waves.

Considering both SU(4) and SU(6) theories allows us to compare how the LLR algorithm performs for both the first-order confinement transition of the former and the much stronger first-order bulk transition of the latter.
Our current results from $12^3 \times 6$ lattices are insufficient to resolve the SU(4) confinement transition, apparently because its small latent heat $\Delta E / V \ll 1$ can only be resolved with much more precise data.
Larger lattice volumes may help to reduce statistical uncertainties, but lead to increased computational costs.
Smaller values of the energy interval size \de may also be necessary, but these would further increase statistical uncertainties, as can be seen from \eq{eq:robmon}.

The comparison between the SU(4) confinement transition and the much stronger SU(6) bulk transition, for which we find $\Delta E / V \approx 3.6$, highlights that the LLR approach is significantly more straightforward for strong first-order phase transitions with large latent heat.
Because this is precisely the situation in which importance sampling can be expected to struggle, our results motivate further studies of density of states approaches focused on strong transitions with large latent heat.
In addition to searching for the SU(6) confinement transition, we therefore plan to expand our studies to SU($N$) lattice Yang--Mills theories with even larger $N > 6$.
We are also exploring broader applications of the LLR algorithm, including to phase transitions in bosonic matrix models of interest in the context of holographic gauge/gravity duality.
% ------------------------------------------------------------------

% ------------------------------------------------------------------
\vspace{20 pt}
\noindent \textsc{Acknowledgments:}~We thank the LSD Collaboration for ongoing joint work investigating composite dark matter and gravitational waves.
We also thank Kurt Langfeld, Paul Rakow, David Mason, James Roscoe and Johann Ostmeyer for helpful conversations about the LLR algorithm.
Numerical calculations were carried out at the University of Liverpool and through the Lawrence Livermore National Laboratory Institutional Computing Grand Challenge program.
DS was supported by UK Research and Innovation Future Leader Fellowship {MR/S015418/1} and STFC grant {ST/T000988/1}.

\bibliographystyle{JHEP}
\bibliography{main}

\providecommand{\href}[2]{#2}\begingroup\raggedright\begin{thebibliography}{10}

\bibitem{Borsanyi:2022xml}
S.~Borsanyi, R.~Kara, Z.~Fodor, D.A.~Godzieba, P.~Parotto and D.~Sexty,
  \emph{{Precision study of the continuum SU(3) Yang-Mills theory: How to use
  parallel tempering to improve on supercritical slowing down for first order
  phase transitions}},
  \href{https://doi.org/10.1103/PhysRevD.105.074513}{\emph{Phys. Rev. D}
  {\bfseries 105} (2022) 074513}
  [\href{https://arxiv.org/abs/2202.05234}{{\ttfamily 2202.05234}}].

\bibitem{Langfeld:2022uda}
K.~Langfeld, P.~Buividovich, P.E.L.~Rakow and J.~Roscoe, \emph{{Reduced
  critical slowing down for statistical physics simulations}},
  \href{https://doi.org/10.1103/PhysRevE.106.054139}{\emph{Phys. Rev. E}
  {\bfseries 106} (2022) 054139}
  [\href{https://arxiv.org/abs/2204.04712}{{\ttfamily 2204.04712}}].

\bibitem{Caprini:2019egz}
C.~Caprini, M.~Chala, G.C.~Dorsch, M.~Hindmarsh, S.J.~Huber, T.~Konstandin
  et~al., \emph{{Detecting gravitational waves from cosmological phase
  transitions with LISA: an update}},
  \href{https://doi.org/10.1088/1475-7516/2020/03/024}{\emph{JCAP} {\bfseries
  2003} (2020) 024} [\href{https://arxiv.org/abs/1910.13125}{{\ttfamily
  1910.13125}}].

\bibitem{LatticeStrongDynamics:2020jwi}
{\scshape LSD} collaboration, \emph{{Stealth dark matter confinement transition
  and gravitational waves}},
  \href{https://doi.org/10.1103/PhysRevD.103.014505}{\emph{Phys. Rev. D}
  {\bfseries 103} (2021) 014505}
  [\href{https://arxiv.org/abs/2006.16429}{{\ttfamily 2006.16429}}].

\bibitem{Huang:2020mso}
W.-C.~Huang, M.~Reichert, F.~Sannino and Z.-W.~Wang, \emph{{Testing the dark
  SU($N$) Yang--Mills theory confined landscape: From the lattice to
  gravitational waves}},
  \href{https://doi.org/10.1103/PhysRevD.104.035005}{\emph{Phys. Rev. D}
  {\bfseries 104} (2021) 035005}
  [\href{https://arxiv.org/abs/2012.11614}{{\ttfamily 2012.11614}}].

\bibitem{Kang:2021epo}
Z.~Kang, S.~Matsuzaki and J.~Zhu, \emph{{Dark confinement-deconfinement phase
  transition: a roadmap from Polyakov loop models to gravitational waves}},
  \href{https://doi.org/10.1007/JHEP09(2021)060}{\emph{JHEP} {\bfseries 2109}
  (2021) 060} [\href{https://arxiv.org/abs/2101.03795}{{\ttfamily
  2101.03795}}].

\bibitem{Appelquist:2015yfa}
{\scshape LSD} collaboration, \emph{{Stealth Dark Matter: Dark scalar baryons
  through the Higgs portal}},
  \href{https://doi.org/10.1103/PhysRevD.92.075030}{\emph{Phys. Rev. D}
  {\bfseries 92} (2015) 075030}
  [\href{https://arxiv.org/abs/1503.04203}{{\ttfamily 1503.04203}}].

\bibitem{Appelquist:2015zfa}
{\scshape LSD} collaboration, \emph{{Detecting Stealth Dark Matter Directly
  through Electromagnetic Polarizability}},
  \href{https://doi.org/10.1103/PhysRevLett.115.171803}{\emph{Phys. Rev. Lett.}
  {\bfseries 115} (2015) 171803}
  [\href{https://arxiv.org/abs/1503.04205}{{\ttfamily 1503.04205}}].

\bibitem{Kribs:2018ilo}
G.D.~Kribs, A.~Martin, B.~Ostdiek and T.~Tong, \emph{{Dark Mesons at the LHC}},
  \href{https://doi.org/10.1007/JHEP07(2019)133}{\emph{JHEP} {\bfseries 1907}
  (2019) 133} [\href{https://arxiv.org/abs/1809.10184}{{\ttfamily
  1809.10184}}].

\bibitem{Butterworth:2021jto}
J.M.~Butterworth, L.~Corpe, X.~Kong, S.~Kulkarni and M.~Thomas, \emph{{New
  sensitivity of LHC measurements to composite dark matter models}},
  \href{https://doi.org/10.1103/PhysRevD.105.015008}{\emph{Phys. Rev. D}
  {\bfseries 105} (2022) 015008}
  [\href{https://arxiv.org/abs/2105.08494}{{\ttfamily 2105.08494}}].

\bibitem{Langfeld:2012ah}
K.~Langfeld, B.~Lucini and A.~Rago, \emph{{The density of states in gauge
  theories}}, \href{https://doi.org/10.1103/PhysRevLett.109.111601}{\emph{Phys.
  Rev. Lett.} {\bfseries 109} (2012) 111601}
  [\href{https://arxiv.org/abs/1204.3243}{{\ttfamily 1204.3243}}].

\bibitem{Langfeld:2015fua}
K.~Langfeld, B.~Lucini, R.~Pellegrini and A.~Rago, \emph{{An efficient
  algorithm for numerical computations of continuous densities of states}},
  \href{https://doi.org/10.1140/epjc/s10052-016-4142-5}{\emph{Eur. Phys. J. C}
  {\bfseries 76} (2016) 306}
  [\href{https://arxiv.org/abs/1509.08391}{{\ttfamily 1509.08391}}].

\bibitem{Korner:2020vjw}
M.~K\"orner, K.~Langfeld, D.~Smith and L.~von Smekal, \emph{{Density of states
  approach to the hexagonal Hubbard model at finite density}},
  \href{https://doi.org/10.1103/PhysRevD.102.054502}{\emph{Phys. Rev. D}
  {\bfseries 102} (2020) 054502}
  [\href{https://arxiv.org/abs/2006.04607}{{\ttfamily 2006.04607}}].

\bibitem{Lucini:2012gg}
B.~Lucini and M.~Panero, \emph{{SU($N$) gauge theories at large $N$}},
  \href{https://doi.org/10.1016/j.physrep.2013.01.001}{\emph{Phys. Rept.}
  {\bfseries 526} (2013) 93} [\href{https://arxiv.org/abs/1210.4997}{{\ttfamily
  1210.4997}}].

\bibitem{Lucini:2005vg}
B.~Lucini, M.~Teper and U.~Wenger, \emph{{Properties of the deconfining phase
  transition in SU(N) gauge theories}},
  \href{https://doi.org/10.1088/1126-6708/2005/02/033}{\emph{JHEP} {\bfseries
  0502} (2005) 033} [\href{https://arxiv.org/abs/hep-lat/0502003}{{\ttfamily
  hep-lat/0502003}}].

\bibitem{Mason:2022trc}
D.~Mason, B.~Lucini, M.~Piai, E.~Rinaldi and D.~Vadacchino, \emph{{The density
  of states method in Yang--Mills theories and first-order phase transitions}},
  \href{https://doi.org/10.1051/epjconf/202227408007}{\emph{EPJ Web Conf.}
  {\bfseries 274} (2022) 08007}
  [\href{https://arxiv.org/abs/2211.10373}{{\ttfamily 2211.10373}}].

\bibitem{Mason:2022aka}
D.~Mason, B.~Lucini, M.~Piai, E.~Rinaldi and D.~Vadacchino, \emph{{The density
  of state method for first-order phase transitions in Yang--Mills theories}},
  \href{https://arxiv.org/abs/2212.01074}{{\ttfamily 2212.01074}}.

\bibitem{Springer:2021liy}
F.~Springer and D.~Schaich, \emph{{Density of states for gravitational waves}},
  \href{https://doi.org/10.22323/1.396.0043}{\emph{Proc. Sci.} {\bfseries
  LATTICE2021} (2022) 043} [\href{https://arxiv.org/abs/2112.11868}{{\ttfamily
  2112.11868}}].

\bibitem{Springer:2022qos}
F.~Springer and D.~Schaich, \emph{{Progress applying density of states for
  gravitational waves}},
  \href{https://doi.org/10.1051/epjconf/202227408008}{\emph{EPJ Web Conf.}
  {\bfseries 274} (2022) 08008}
  [\href{https://arxiv.org/abs/2212.09199}{{\ttfamily 2212.09199}}].

\bibitem{Langfeld:2016kty}
K.~Langfeld, \emph{{Density-of-states}},
  \href{https://doi.org/10.22323/1.256.0010}{\emph{Proc. Sci.} {\bfseries
  LATTICE2016} (2017) 010} [\href{https://arxiv.org/abs/1610.09856}{{\ttfamily
  1610.09856}}].

\bibitem{Creutz:1987xi}
M.~Creutz, \emph{{Overrelaxation and Monte Carlo Simulation}},
  \href{https://doi.org/10.1103/PhysRevD.36.515}{\emph{Phys. Rev. D} {\bfseries
  36} (1987) 515}.

\bibitem{Katznelson:1984kw}
E.~Katznelson and A.~Nobile, \emph{{Implementation and Statistical Analysis of
  Metropolis Algorithm for SU(3)}},
  \href{https://doi.org/10.1016/0010-4655(86)90159-1}{\emph{Comput. Phys.
  Commun.} {\bfseries 39} (1986) 1}.

\bibitem{Wingate:2000bb}
M.~Wingate and S.~Ohta, \emph{{Deconfinement transition and string tensions in
  SU(4) Yang--Mills theory}},
  \href{https://doi.org/10.1103/PhysRevD.63.094502}{\emph{Phys. Rev. D}
  {\bfseries 63} (2001) 094502}
  [\href{https://arxiv.org/abs/hep-lat/0006016}{{\ttfamily hep-lat/0006016}}].

\bibitem{Francesconi:2019nph}
O.~Francesconi, M.~Holzmann, B.~Lucini and A.~Rago, \emph{{Free energy of the
  self-interacting relativistic lattice Bose gas at finite density}},
  \href{https://doi.org/10.1103/PhysRevD.101.014504}{\emph{Phys. Rev. D}
  {\bfseries 101} (2020) 014504}
  [\href{https://arxiv.org/abs/1910.11026}{{\ttfamily 1910.11026}}].

\bibitem{Francesconi:2019aet}
O.~Francesconi, M.~Holzmann, B.~Lucini, A.~Rago and J.~Rantaharju,
  \emph{{Computing general observables in lattice models with complex
  actions}}, \href{https://doi.org/10.22323/1.363.0200}{\emph{Proc. Sci.}
  {\bfseries LATTICE2019} (2019) 200}
  [\href{https://arxiv.org/abs/1912.04190}{{\ttfamily 1912.04190}}].

\end{thebibliography}\endgroup
\end{document}